\documentclass{emulateapj}
\usepackage{natbib, aas_macros}
\begin{document}

\title{Constraints on the faint end of the quasar luminosity function at $z \sim$ 5 in the COSMOS field}
\author{H. Ikeda,\altaffilmark{1,2} T. Nagao,\altaffilmark{2,3} K. Matsuoka,\altaffilmark{1,2,4} Y. Taniguchi,\altaffilmark{5} Y. Shioya,\altaffilmark{5} M. Kajisawa,\altaffilmark{1,5} M. Enoki,\altaffilmark{6} P. Capak,\altaffilmark{7} F. Civano,\altaffilmark{8} A. M. Koekemoer,\altaffilmark{9} D. Masters,\altaffilmark{7,10} T. Morokuma,\altaffilmark{11} M. Salvato,\altaffilmark{12} E. Schinnerer,\altaffilmark{13} and N. Z. Scoville\altaffilmark{7} }
\altaffiltext{1}{Graduate School of Science and Engineering, Ehime University, Bunkyo-cho, Matsuyama 790-8577, Japan; email: ikeda@cosmos.phys.sci.ehime-u.ac.jp}
\altaffiltext{2}{Department of Astronomy, Graduate School of Science, Kyoto University, Kitashirakawa-Oiwake-cho, Sakyo-ku, Kyoto 606-8502,
Japan}
\altaffiltext{3}{The Hakubi Center for Advanced Research, Kyoto University, Yoshida-Ushinomiya-cho, Sakyo-ku, Kyoto 606-8302, Japan}
\altaffiltext{4}{Research Fellow of the Japan Society for the Promotion of Science}
\altaffiltext{5}{Research Center for Space and Cosmic Evolution, Ehime University, Bunkyo-cho, Matsuyama 790-8577, Japan}
\altaffiltext{6}{Faculty of Bussiness Administration, Tokyo Keizai University, 1-7-34 Minami-cho, Kokubunji, Tokyo 185-8502, Japan}
\altaffiltext{7}{California Institute of Technology, MC 105-24, 1200 East California Boulevard, Pasadena, CA 91125, USA}
 \altaffiltext{8}{Harvard Smithsonian Center for Astrophysics, 60 Garden St., Cambridge, MA 02138, USA}
 \altaffiltext{9}{Space Telescope Science Institute, 3700 San Martin Drive, Baltimore, MD 21218, USA}
\altaffiltext{10}{Department of Physics and Astronomy, University of California, 900 University Ave, Riverside, CA 92521, USA}
 \altaffiltext{11}{Institute of Astronomy, Graduate School of Science, University of Tokyo, 2-21-1 Osawa, Mitaka 181-0015, Japan}
 \altaffiltext{12}{Max-Planck-Institut f\"ur Plasmaphysik, Boltzmanstrasse 2, D-85741 Garching, Germany}
\altaffiltext{13}{Max-Planck-Institut f\"ur Astronomie, K\"onigstuhl 17, D-69117 Heidelberg, Germany}
\shortauthors{Ikeda et al.}    

\begin{abstract}
 We present the result of our low-luminosity quasar survey in the redshift range of
 4.5 $\lesssim$  {\it z} $\lesssim$  5.5 in the COSMOS field. Using the COSMOS 
 photometric catalog, we selected 15 quasar candidates with 22
\verb|<|  {\it i\arcmin} \verb|<|  24 at {\it z} $\sim$\ 5, that are $\sim$ 3 mag
fainter than the SDSS quasars in the same redshift range. 
We obtained optical spectra for 14 of the 15 candidates using FOCAS on the Subaru 
Telescope and did not identify any low-luminosity type-1 quasars at $z\sim5$ while a low-luminosity type-2 quasar at $z\sim5.07$ was discovered. 
In order to constrain the faint end of the quasar luminosity function at $z\sim5$,
 we calculated the 1$\sigma$ confidence upper limits of the space density of type-1 quasars. As a result, 
 the 1$\sigma$ confidence upper limits on the quasar space density are $\Phi<$ 1.33 $\times$ 10$^{-7}$ Mpc$^{-3}$ mag$^{-1}$ for 
$-24.52 < M_{1450} < -23.52$ and  $\Phi<$ 2.88 $\times$ 10$^{-7}$ Mpc$^{-3}$ mag$^{-1}$ for $-23.52 < M_{1450} < -22.52$. 
The inferred 1$\sigma$ confidence upper limits
of the space density are then used to provide constrains on
the faint-end slope and the break absolute magnitude of the quasar
luminosity function at $z\sim5$.
We find that the quasar space density decreases gradually as a function of redshift at low luminosity 
($M_{1450}\sim -23$), being similar to the trend found for quasars 
with high luminosity ($M_{1450}<-26$). This result is consistent with the so-called
downsizing evolution of quasars seen at lower redshifts. \end{abstract}

\newpage

\section{Introduction}
The evolution of supermassive black holes (SMBHs) is now
regarded as one of the most important unresolved issues
in the modern astronomy, after the discovery of the
galaxy-SMBH ``co-evolution" inferred from, e.g., a tight
relationship between the mass of SMBHs and their host
bulges \citep[e.g.,][]{2003ApJ...589L..21M, 2004ApJ...604L..89H, 2009ApJ...698..198G}.
Measuring the whole shape of the quasar luminosity function
(QLF) is particularly important to understand how the SMBHs
grew, since it is highly dependent on some key parameters
of SMBHs such as the growth timescale of SMBHs \citep[e.g.,][]{
2003PASJ...55..133E}. 

The QLF at $z\lesssim3$ has been well quantified over a wide luminosity range \citep[e.g.,][]{2009MNRAS.399.1755C} 
and is best represented by a double power law \citep[e.g.,][]{1988MNRAS.235..935B, 1995ApJ...438..623P}. 
Recently, the faint end of the QLF at $z\sim4$ has been measured \citep{2010ApJ...710.1498G, 2011ApJ...728L..25I, 2011ApJ...728L..26G} and is also best represented by 
a double power law. More interestingly, recent studies on the 
optical QLF show that the space density of low-luminosity active galactic nuclei (AGNs) peaks at a lower redshift 
than that of more luminous AGNs \citep{2009MNRAS.399.1755C, 2011ApJ...728L..25I}. 
This result can be interpreted 
as AGN  (or  SMBH) downsizing evolution, since the brighter AGNs tend to harbor the more massive SMBHs if the dispersion of the Eddington ratio of quasars
is not very large \citep[see, e.g.,][]{2011ApJ...733...60T}.
The AGN downsizing has been also reported by X-ray surveys (\citealt{2003ApJ...598..886U}; \citealt{2005A&A...441..417H}; see also \citealt{2009ApJ...693....8B} and \citealt{2011ApJ...741...91C}). However, the physical origin of the AGN downsizing is totally unclear, that makes high-$z$ low-luminosity
quasar surveys more important \citep[see][for theoretical works on the AGN downsizing evolution]{2012MNRAS.419.2797F}.

Recently, some low-luminosity quasar surveys have been performed at $z>5$ \citep{2006AJ....132..823C,2005ApJ...634L...9M}.
\cite{2006AJ....132..823C} identified three quasars at $z>5$ with $z'<22$ and included a quasar at $z$ = 5.85 with $z'$ = 20.68, 
in the AGES survey which covers 8.5 $\rm deg^2$. 
Jiang et al. (2008) also identified five new quasars at $z>5.8$ with $20 < z' < 21$ in the Sloan Digital Sky Survey (SDSS) deep stripe which covers 260 $\rm deg^2$.
The space density of quasars at $z\sim6$ which is calculated by the result of Cool et al. (2006) is about six times larger than the result of Jiang et al. (2008). This large discrepancy may be caused by the small survey area of Cool et al. (2006). \cite{2005ApJ...634L...9M} identified a very faint quasar at $z=5.70$ with $z' =$ 23.0 in the total quasar survey area of 2.5 $\rm deg^2$. \cite{2005ApJ...634L...9M} mentioned that the surface density of quasars at similar redshifts is roughly consistent with previous extrapolations of the faint end of the QLF. In this way, some low-luminosity quasars have been discovered although the faint end of QLF is not determined exactly, due to the lack of low-luminosity quasars.

At $z>6$, a number of luminous quasars have been found up to $z\sim6.5$ by the SDSS \citep[e.g.,][]{2006AJ....131.1203F, 2006MNRAS.371..769G, 2008AJ....135.1057J, 2009AJ....138..305J} and the Canada-France High-$z$ Quasar Survey (CFHQS; \citealt{2007AJ....134.2435W}; \citealt{2009AJ....137.3541W}; \citealt{2010AJ....139..906W}). Recently, a luminous quasar at $z=7.085$ has been found \citep{2011Natur.474..616M} through the data obtained by the United Kingdom Infrared Telescope Infrared Deep Sky Survey \citep[UKIDSS;][]{2007MNRAS.379.1599L}.
Although some low-luminosity quasars at $z>5$ have been discovered as mentioned above, the faint-end slope of the $z>5$ QLF is still very poorly constrained. 
Consequently it is not understood how low-luminosity quasars evolve at high redshifts, or if the AGN downsizing evolution is also seen in the earlier universe. 
Since the number density of low-luminosity quasars is expected to be much higher than that of high-luminosity quasars, the whole picture of SMBH evolution cannot be understood without firm measurements of the number density of low-luminosity quasars at such high redshifts.

Motivated by these issues, we have searched for low-luminosity quasars at $z\sim5$ in the COSMOS field \citep{2007ApJS..172....1S}.
In Section 2, we describe the data and method that were used for the photometric selection of quasar candidates.
In Section 3, we report the results of the follow-up spectroscopic observations. 
In Section 4, we describe how we estimate the photometric completeness to derive the QLF. In Section 5, we present the upper limits of the QLF at $z\sim5$ and briefly discuss it. Throughout this paper we use a $\Lambda$ cosmology with $\Omega_m$ = 0.3,\  $\Omega_{\Lambda}$ = 0.7, and the Hubble constant of $H_0$ = 70 km s$^{-1}$ Mpc$^{-1}$.\\

\section{The Sample}

\subsection{The Cosmic Evolution Survey}
The COSMOS is a treasury program on the Hubble Space Telescope (HST) and comprises 270 and 320 orbits allocated with HST Cycles 12 and 13, respectively (Scoville et al. 2007; Koekemoer et al. 2007).
The COSMOS field covers an area of $\rm 1.4^{\circ} \times 1.4^{\circ}$ square which corresponds to $2\deg^2$, centered at R.A. (J2000) = 10:00:28.6 and Dec. (J2000) =+02:12:21.0.

We use the official COSMOS photometric redshift catalog for photometry (\citealt{2009ApJ...690.1236I}; see also \citealt{2007ApJS..172...99C}) to select the quasar photometric candidates at $z\sim5$. 
This catalog covers an area of $\sim$ 2 $\rm deg^2$ and contains several 
photometric measurements. Specifically in this paper, we use the $u^*$-band $3^{''}$ diameter aperture apparent magnitude measured on the image obtained with MegaCam \citep{2003SPIE.4841...72B} on the Canada-France-Hawaii Telescope (CFHT), and the $3^{''}$ diameter aperture apparent magnitudes of the $g'$-, $r'$-, $i'$-, and $z'$-bands  \citep{2007ApJS..172....9T}
 measured on the image obtained with the Subaru Suprime-Cam \citep{2002PASJ...54..833M}, and the $i'$-band total apparent magnitude\citep[${\tt MAG\_AUTO}$ measured by SExtractor;][]{1996A&AS..117..393B} whose measurement is also based on the Suprime-Cam i'-band image.

\begin{figure}[t!]
\begin{center}
\includegraphics[bb=0 0 500 450,clip,width=8cm]{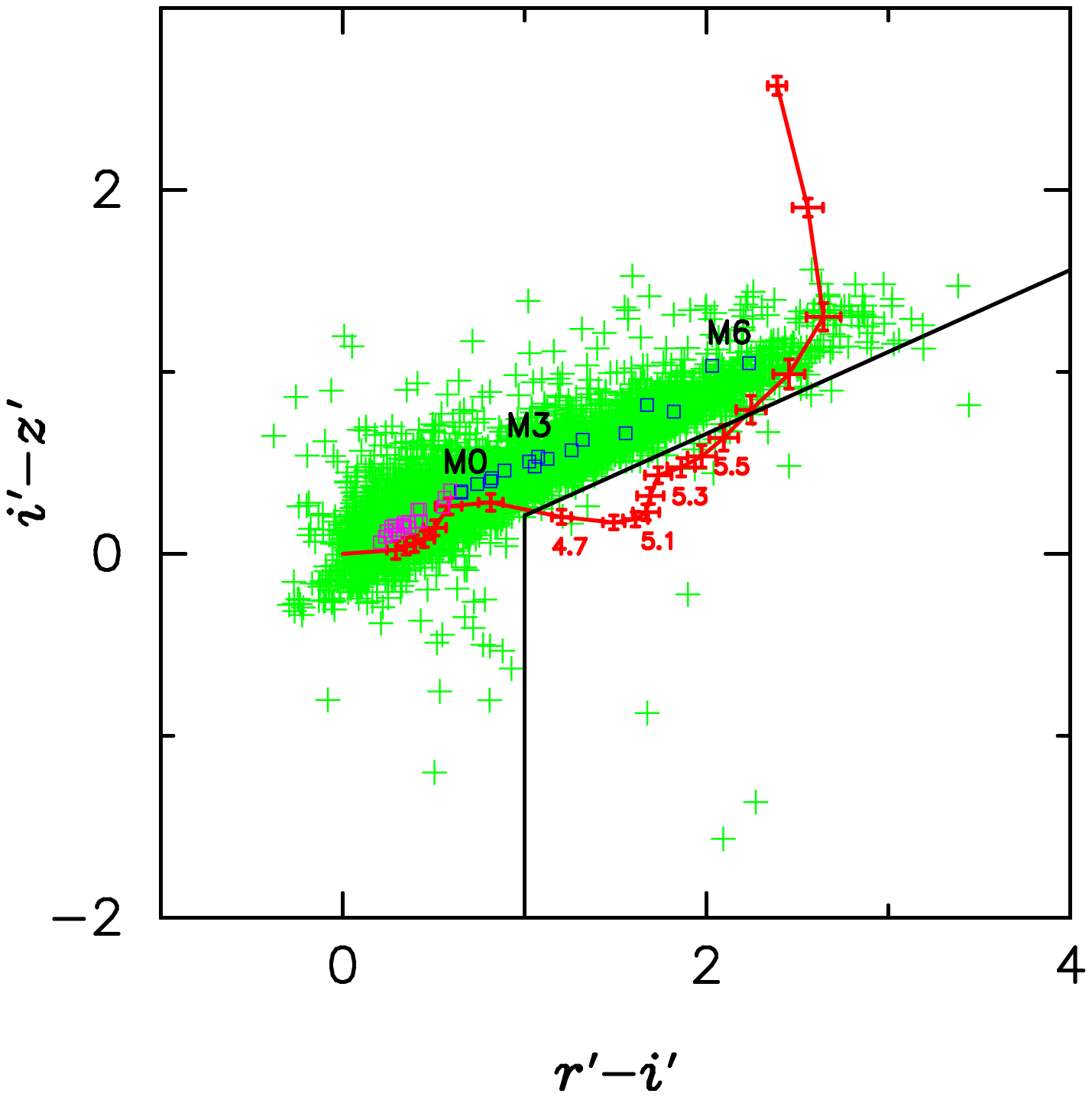}

\end{center}
\caption{Two color diagram of $r'-i'$ and $i'-z'$, that is used for our quasar selection. Green plus symbols denote point sources with $22<i'({\tt MAG\_AUTO})<24$. Blue and magenta squares denote colors of M-type stars and  K-type stars \citep{1998PASP..110..863P}, respectively. The  red line is the median track of the model quasar colors. The red error bars show the standard deviation of the $r'-i'$ and $i'-z'$ colors in our model quasar spectra. The black solid line shows our photometric criteria used to select quasar candidates at $z\sim5$.}

\label{fig:Two color diagram used in our quasar selection.}

\end{figure}

The 5$\sigma$ limiting AB apparent magnitudes are $u^*$ = 26.5, $g'$ = 26.5, $r'$ = 26.6, $i'$ = 26.1, and $z'$ = 25.1 ($3^{''}$ diameter aperture). 
Since we also use the Advanced Camera for Surveys (ACS) catalog \citep{2007ApJS..172..196K, 2007ApJS..172..219L}
to separate galaxies from point sources, our survey area is restricted to the area mapped with ACS on HST ($1.64\ \rm deg^2$).
Note that all of the data in the official COSMOS 
photometric redshift catalog overlaps the entire ACS field.
\subsection{Quasar Candidate Selection}
\begin{table*}
\begin{center}
\tabcaption{Properties of the quasar candidates at $z\sim5$}

\begin{tabular}{ccc@{\hspace{0.5cm}}c@{\hspace{0.5cm}}c@{\hspace{0.5cm}}c@{\hspace{0.5cm}}c@{\hspace{0.5cm}}c@{\hspace{0.5cm}}c@{\hspace{0.5cm}}c} \hline
\tableline\tableline\noalign{\smallskip}
 Number &  & &R.A.&  Decl. & $i'$ ({\tt MAG\_AUTO})&  $r'-i'$ &  $i'-z'$& $\rm Exp. Time^{a}$ \\
\noalign{\smallskip}\tableline\noalign{\smallskip}
   &&& (deg) &(deg) & (mag)& (mag) & (mag)&(sec)& \\ \hline
        1 & && 150.69131 & 1.637161 & 23.40 & 2.34  &0.67 &2400\\
        2 & && 150.45275 & 1.669653 & 23.48  & 2.04&2.69    &3000\\
        3 & && 150.17448 & 1.629074 & 23.76  & 1.67 &2.11&1800\\ 
        4$^{b}$ & && 150.64917 & 1.816186 & 23.39  & 3.44&0.82&-\\
        5 &  && 149.87082 & 1.882791 & 23.98  & 1.26&0.17 &2400\\
        6 &  &&149.85403 & 1.823611   & 23.97  & 2.58&0.87&2400\\
        7 &  &&149.78245 & 2.221621 & 23.96 & 3.19&1.13 &1800\\
        8 &  &&149.69804 & 2.283260 & 23.67  & 2.04&0.44&1800 \\ 
        9 &  &&150.56861 & 2.317432 & 23.98 & 4.09&1.26&1800\\ 
        10 &  &&150.05481 & 2.376726 & 23.89 & 1.09&0.25& 2700\\
         $11^{c}$ &  &&149.78381 & 2.452135 & 23.70  & 1.35&0.26&7200\\
         12 &  &&150.16401 & 2.549605 & 23.31  & 1.96&0.61&2400 \\
          13 &  &&149.96443 & 2.473646 & 23.93  & 1.21&0.21 &2700\\
          14 &  &&150.66035 & 2.786445 & 23.51 & 1.93& 0.57&2400\\
           15 &  &&149.59161 & 2.659749 & 23.16  & 2.26&0.77&  2400\\ \hline
\noalign{\smallskip}\tableline\noalign{\smallskip}
\end{tabular}\flushleft{$^{a}$Total on-source exposure time in the FOCAS spectroscopic observation.\\
        $^{b}$ This quasar candidate was not observed.\\
        $^{c}$ Type-2 quasar at $z=5.07$.}  
\end{center}

\end{table*}
Quasars at $z\sim5$ show the Lyman break in their spectral energy distribution (SED) that 
falls between the wavelengths of  the $r'$ and $i'$ filters, making their $r'-i'$ color redder than their $i'-z'$ color. 
We utilize this property to select candidates of low-luminosity quasars at $z\sim5$. 
Here typical quasar colors as a function of redshift are necessary to 
define reliable color-selection criteria for quasars at $z \sim 5$. 
Therefore we generate  model quasar spectra following the procedure 
generally adopted in previous studies \citep[e.g.,][]{1999AJ....117.2528F, 2004ApJ...605..625H, 2006AJ....131.2766R, 2008ApJ...675...49S}, 
and derive the $g'-r'$, $r'-i'$, and $i'-z'$ colors of the model quasars
at redshifts from 0 to 6.
In this procedure, we adopt the typical power-law slope ($\alpha_{\nu}$ = 0.46, where $f_{\nu}\propto\nu^{-\alpha_{\nu}}$), 
 Ly$\alpha$ rest-frame equivalent width ($\it EW_{\rm 0}$ = 90 $\rm$\AA), and typical emission-line flux ratios \citep{2001AJ....122..549V}. 
 The effects of the intergalactic absorption by the neutral hydrogen were corrected by adopting the extinction model of \cite{1995ApJ...441...18M}.
Our simulated colors of the model quasars are shown in the $r'-i'$ versus $i'-z'$ diagram (Fig. 1). 
Note that the dispersions in the
power-law slope and EWs of quasars are also taken into account
when we calculate the photometric completeness of our survey (see Section 4).

We select our quasar photometric candidates at $z \sim 5$ using the following criteria:
 \begin{eqnarray}
  22<i' {\rm ({\tt MAG\_AUTO})}<24, \\
  i'-z'<0.45(r'-i')-0.24,  \end{eqnarray}
   \begin{equation}
 u^* > 27.49,   \end{equation}
 \begin{equation}
 g'-r' \geq 1.3,   \end{equation}
 {\rm and}  
 \begin{equation}
   r'-i'>1.0.
 \end{equation}
The criterion (2) is used to select quasars efficiently without significant contamination from stars (especially from M0 to M6, see also Fig. 1), taking the color distribution of stars and model quasars into account. 
 To remove possible foreground contaminations further, we introduce the additional criteria (3), (4), and (5). 
 These latter two color thresholds are adopted by taking empirical color
distributions of quasars at $z\sim5$ into account \citep{2006AJ....131.2766R}. 

Here we comment on our point-source criterion based on the HST image (F814W, see Koekemoer et al. 2007), whose spatial resolution is $0\farcs 09$ in FWHM (that corresponds to $\sim0.6\rm kpc$ at $z=5$). Since the size of high-$z$ quasar host galaxies is larger than 1 kpc \citep{2012MNRAS.420.3621T}, it is expected that the quasar host galaxies  are spatially resolved in the ACS images. Therefore it would be possible some quasars could be excluded from the sample of our quasar photometric candidates.
However, at $z\sim1$, some earlier works show that the host galaxy of quasars with a
similar absolute magnitude to our targets is typically $\sim2$ mag fainter than
their nucleus \citep[e.g.,][]{2001ApJ...546..782M, 2009ApJ...704..415M}.
The brightness difference may be even larger at higher redshifts,
because the typical Eddington ratio of luminous quasars is roughly
constant at $z\sim1-4$ \citep{2009ApJ...700...49T} while the mass ratio of SMBHs
to host galaxies is higher at higher redshifts \citep[e.g.,][]{2006ApJ...640..662T, 
2008ApJ...681..925W, 2006ApJ...645..900W, 2010ApJ...708..137M, 2010ApJ...708.1507B, 2011ApJ...742..107B}.
Therefore we conclude that quasars explored in this work should be recognized as
point sources in the HST image.
To distinguish the galaxies and point sources, Leauthaud et al. (2007)
used the SExtractor parameter $\rm \tt MU\_MAX$ (peak surface brightness
above the background level) and $\rm \tt MAG\_AUTO$ (see Fig. 5 of Leauthaud et al. 2007). Because the fact that the light distribution of a point source varies
with magnitude. Therefore we can distinguish the extended objects and 
point sources by using the $\rm \tt MU\_MAX$ and the $\rm \tt MAG\_AUTO$.

Accordingly we removed 23 spatially-extended objects satisfying the criteria (1)--(5) based on the classification by Leauthaud et al. (2007). 
As a result, we obtain a sample of 15 quasar candidates among 7318 point sources 
with $22 < i'(\tt MAG\_AUTO) < \rm 24$. 
The selected candidates are listed in Table 1.

%
%

\begin{figure}
\begin{center}
\includegraphics[bb=0 10 550 800,clip,width=6cm,angle=270]{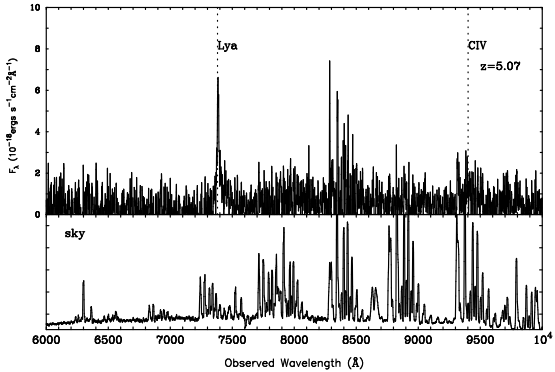}

\end{center}
\caption{Reduced spectrum of object No. 11 in Table 1 (upper panel) and typical sky spectrum (lower panel). Dotted lines show the expected wavelengths of quasar emission lines: Ly$\alpha$ $\lambda1216$ and {C {\sc iv} $\lambda1549$}.}
\label{ quasar spectrum}

\end{figure}

\section{Spectroscopic Observation}
The spectroscopic follow-up observations of the quasar candidates were carried out 
at the Subaru Telescope with  the Faint Object Camera and Spectrograph \citep[FOCAS;][]{2002PASJ...54..819K} on 7--11 January 2010. 
We used the 300 grating with the SO58 filter, whose wavelength coverage is 
$\rm$ 5800\AA  \ $\le$ $\rm \lambda_{obs}$  $\le$ $\rm$10000\AA. 
We used a $0 \farcs 8$-width slit, resulting in a wavelength resolution of 
$R\sim700$ ($\Delta v$ $\sim430$ \rm km $\rm s^{-1}$) as measured by night sky emission lines. 
The typical seeing size was $\sim0 \farcs 7$. Due to the limited observing time, 
we observed 14 of the 15 candidates;
the object No. 4 in Table 1 was not observed.
The individual exposure time was 600 -- 900 sec, and the total exposure time was 1800 -- 7200 sec for each object.

Standard data reduction procedures were performed by using IRAF. 
After the sky subtraction,
we extracted one-dimensional spectra with an aperture size of $1 \farcs 8$ and the relative sensitivity calibration was performed using the spectral data of a spectrophotometric standard star, Feige 34. 
The spectra of 14 objects were then flux-calibrated using the $i'$-band photometric magnitude of these objects. 
We found that one spectrum shows only narrow Ly$\alpha$ emission lines at 
$\lambda_{\rm obs}=7381$$\rm$ \AA ($z\sim5.07$ and $\Delta v_{\rm FWHM}\sim 800$ km $\rm s^{-1}$) without 
any high-ionization lines such as {C {\sc iv} (Fig. 2). Since this object is detected in the X-ray band by the Chandra-COSMOS survey  \citep[][CID-2220]{2009ApJS..184..158E} and its X-ray luminosity is 
3$\times$10$^{44}$ erg/s in the 2-10 keV rest frame band \citep{2011ApJ...741...91C, 2012arXiv1205.5030C}, it can be classified as an AGN. 
We can classify this object as a Type 2 AGN based on  the upper limit available for the 
X-ray hardness ratio consistent with mild obscuration (NH$<$5$\times$10$^{23}$cm$^{-2}$) together with its Ly $\alpha$ spectral profile. 
Although the X-ray hardness ratio is not available for this object, we classify
this object as a type-2 AGN based on its Ly$\alpha$ spectral profile.
Therefore we conclude that no type-1 quasars are identified in our spectroscopic follow-up campaign.
The spectra of the remaining 13 objects are consistent with those of Galactic
late-type stars, and an example of these spectra is shown in Fig. 3. 

\section{Completeness}
Quasar surveys are generally not perfectly complete due to various factors such as photometric errors and intrinsic variations in the spectra. Therefore, to derive the QLF acculately, the survey completeness needs to be estimated as a function of the quasar redshift and apparent magnitude. We derive the completeness by modeling quasar spectra, in a similar way as described in Section 2.2.
Here we also take into account the intrinsic variation in the continuum slope and EWs of the emission lines. We assume a Gaussian distribution of the power-law slope $\alpha_{\nu}$ ($f_{\nu}\propto\nu^{-\alpha_{\nu}}$) and Ly$\alpha$ EWs, with means of 0.46 and 90{\AA} (the same as those in Section 2.2), and a standard deviation of 0.30 and 20{\AA}, respectively \citep{2001AJ....122..549V, 1996PASA...13..212F, 2004ApJ...605..625H,2010ApJ...710.1498G}. 
We include emission lines whose flux is larger than $0.5\%$ of the Ly$\alpha$ flux, given in Table 2 of Vanden Berk et al. (2001). The emission-line ratios are assumed to be the same for  all model quasars (i.e., scaling to the Ly$\alpha$ strength). We also include the Balmer continuum and Fe {\sc ii} features by using the template of \cite{1996ApJ...470L..85K}. We create 1000 quasar spectra at each $\Delta$$z$ = 0.01 in the redshift range of $0<z<6$. The effects of intergalactic absorption by neutral hydrogen were corrected by adopting the extinction model of Madau (1995). Then, we calculated the colors of the model quasars in the observed frame.
We compared the colors of the simulated quasars with the empirical quasar colors (SDSS DR7) to check whether the simulated quasar colors are consistent with the empirical quasar colors. 
Fig. 4 shows the comparison between the empirical and simulated quasar colors of $g^*-r^*$, $r^*-i^*$, and $i^*-z^*$ by using the transmission curves of the SDSS filters. 
Note that the dispersion of the simulated quasar colors is systematically smaller than the dispersion of the observed quasar colors, through the average color is consistent between the simulated and observed quasar colors. This is because the simulated colors presented here do not take photometric errors into account. The photometric errors are properly taken into account when the completeness is  calculated, as described below. 

\begin{figure}
\begin{center}
\includegraphics[bb=0 10 550 800,clip,width=6cm,angle=270]{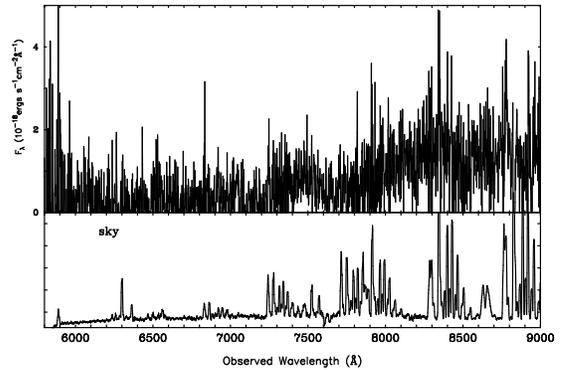}

\end{center}
\caption{Example of the late type star spectrum (upper panel; No. 2 in Table 1) and typical sky spectrum (lower panel). \\}
\label{ quasar spectrum}

\end{figure}

\begin{figure}
\begin{center}
\includegraphics[bb=0 0 550 750,clip,width=7cm]{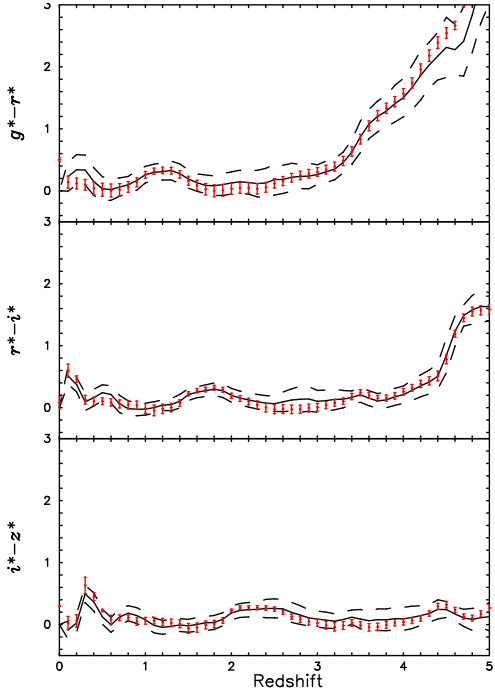}

\end{center}
\caption{Comparison between the empirical and simulated quasar colors. Solid black lines show the empirical mean colors of the SDSS DR7 quasars. Dashed black lines show 1 sigma dispersions of the SDSS DR7 quasar colors. Red points show the simulated quasar colors. Red vertical lines show the standard deviation of simulated quasar colors which do not include photometric errors.}
\label{ quasar spectrum}

\end{figure}

To estimate the photometric completeness, we put the 1000 model quasars at each grid point in apparent magnitude and redshift into Subaru Suprime-Cam FITS images as point sources, using the IRAF {\tt mkobjects} task in the {\tt artdata} package. As for 1000 model quasars, they were generated on a Monte Carlo method of drawing a value of alpha and EW. These point sources have apparent magnitudes calculated from their simulated spectra, in each image ($g'$, $r'$, $i'$, and $z'$). We then tried to detect them in the $i'$-band image with SExtractor, and measure their colors in the double-image mode. 
Note that the measured apparent magnitudes and colors of the simulated quasars are generally different from the magnitudes and colors before inserted into the Suprime-Cam images, due to the effects of photometric errors and neighboring foreground objects. Accordingly, some model quasars are not selected as photometric candidates of quasar with the criteria (1) -- (5). To calculate the fraction of model quasars that are selected as photometric candidates in the above process, we estimate the photometric completeness at various magnitudes and redshifts (Fig. 5). The redshift range of the inferred completeness is moderately high at 4.5 $\lesssim$ {\it z} $\lesssim$ 5.5. 
More specifically, the inferred completeness is $\sim 90\%$ for quasars with $i' = 22.5$, and $\sim80\%$ for those with $i' = 23.5$ in that redshift range.	
The small dip at $z\sim5.2-5.3$ in the estimated completeness is due to the {C {\sc iv} emission that causes the $i'-z'$ color to be red at that redshift range.

\section{Quasar Luminosity Function}
To calculate the upper limits of the quasar space density, we compute the 
effective comoving volume of the survey as:\\
\begin{eqnarray}
	V_{\rm eff} (m_{i'})= d \Omega \int_{z = 0}^{z = \infty}C(m_{i'}, z)\frac{dV}{dz}dz,  
	\end{eqnarray}where $d\Omega$ is the solid angle of the survey and $C(m_{i'}, z)$ is the photometric completeness derived in Section 4. 
For comparison with other works, we convert the $i'$-band apparent magnitude to the absolute AB magnitude at 1450\AA  \
  (e.g., Richards et al. 2006; Croom et al. 2009; Glikman et al. 2010):\\
\begin{eqnarray} 
	M_{1450} = m_{i'}+5-5{\rm log}d_L(z)+2.5(1-\alpha_{\nu})\nonumber \\{\rm log}(1+z)+2.5 \alpha_{\nu} {\rm log} \left( \frac{\lambda_{i^{\prime}}}{1450 \rm \AA} \right), 
	\end{eqnarray}
	where $d_L(z)$, $\alpha_{\nu}$, and $\lambda_{i^{\prime}}$ are the luminosity distance, spectral index of the quasar continuum ($f_{\nu}\propto\nu^{-\alpha_{\nu}}$), and the effective wavelength of the $i'$-band ($\lambda_{i^{\prime}}$=7684$\rm \AA$), respectively. We assumed the $\alpha_{\nu}=0.46$ when we used the equation (7). As for the quasar candidates which did not perform the spectroscopic observations, we calculated the absolute magnitude at 1450$\rm \AA$ by assuming the effective redshift. Given the effective comoving volume, the 1 $\sigma$ confidence upper limits on the space density of type-1 quasars are calculated using statistics from \cite{1986ApJ...303..336G}.

\begin{figure}[t]
\begin{center}
\includegraphics[bb=0 100 600 450,clip,width=9.5cm]{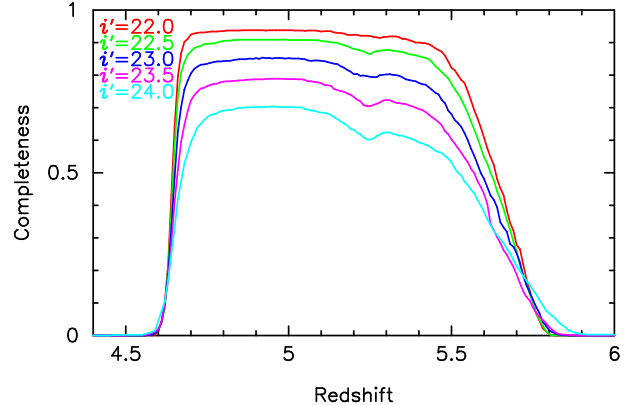}

\end{center}
\caption{Simulated photometric completeness. Red, green, blue, purple, and cyan lines show the completeness for quasars with $m_{i'}$ = 22.0, 22.5, 23.0, 23.5, and 24.0, respectively.}

\label{fig:Completeness}

\end{figure}

The derived 1 $\sigma$ confidence upper limits on the space density of type-1 quasars are $\Phi<$ 1.33 $\times$ 10$^{-7}$ Mpc$^{-3}$ mag$^{-1}$ for 
$-24.52 < M_{1450} < -23.52$ and  $\Phi<$ 2.88 $\times$ 10$^{-7}$ Mpc$^{-3}$ mag$^{-1}$ for $-23.52 < M_{1450} < -22.52$. 
Note that there is another quasar candidate in the magnitude bin of $-23.52 < M_{1450} < -22.52$ which was not observed with FOCAS. 
We take into account the possibility that this candidate
is a quasar when calculating the $1 \sigma$ confidence upper limit on 
the space density of type-1 quasars for $-23.52 < M_{1450} < -22.52$. The obtained 1 $\sigma$ confidence upper limits on the space density of type-1 quasars are plotted in Fig. 6.

To compare our upper limits on the quasar space density with the previous quasar surveys at similar redshifts, we also plot the results of COMBO-17 (Wolf et al. 2003), 
SDSS (Richards et al. 2006), and GOODS \citep{2007A&A...461...39F}, in the redshift ranges of 4.2 $<$  ${\it z}$ $<$ 4.8, 4.5 $<$  ${\it z}$ $<$ 5.0, 
and 4.0 $<$  ${\it z}$ $<$ 5.2 respectively, in Fig. 6. 
Note that the low-luminosity quasar sample of Fontanot et al. (2007) includes type-2 quasars while our sample and the sample of Wolf et al. (2003) do not include type-2 quasars. To compare the result of Fontanot et al. (2007), we also calculated the quasar space density when we included a type-2 quasar and the obtained quasar space density and its error are  $\Phi=$ $0.87_{-0.72}^{+2.01}$ $\times$ 10$^{-7}$ Mpc$^{-3}$ mag$^{-1}$ for $-23.52 < M_{1450} < -22.52$ (see Fig. 6).
This figure shows a marginal discrepancy between the results of Fontanot et al. (2007) and of ours.
However, the redshift difference between the two studies should be taken into account for such a comparison because the quasar space density shows significant redshift evolution. 

 \vspace{0.5cm}
\centerline{{\vbox{\epsfxsize=7.5cm\epsfbox{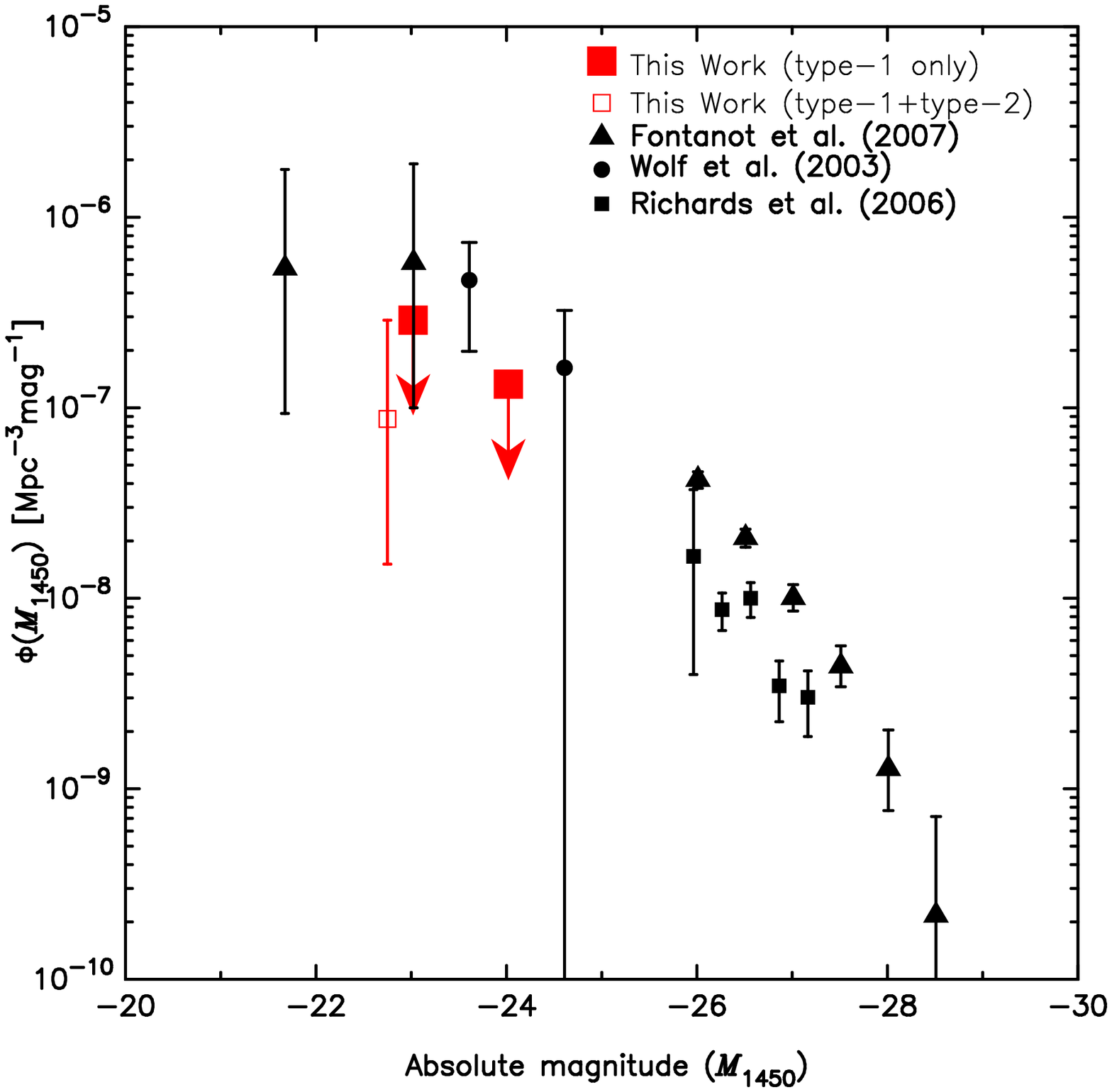}}}}
\figcaption{The $z\sim5$ quasar luminosity functions. The red filed squares show our results (1$\sigma$ confidence limits on the quasar space density) and the red open square shows the quasar space density when we include a type-2 quasar at $z\sim5.07$. For clarifying the data plots in the figure, the open red
square is slightly shifted to the left direction to avoid the overlap of the error bars. Triangles, squares, and circles denote the results reported by Fontanot et al. (2007), Richards et al. (2006), and Wolf et al. (2003), respectively.}
\vspace{0.5cm}

In Fig. 7, we plot the QLF
reported by Fontanot et al. (2007) after correcting for the redshift difference (i.e., taking the redshift evolution in the QLF into account), adopting the model 13a in Fontanot et al. (2007). 
The model 13a assumes a pure density evolution of the QLF with an exponential form, that gives the minimum $\chi^{2}$ among the models examined in Fontanot et al. (2007).
More specifically, in the model 13a, the bright-end slope is fixed to be 3.31 
and there are three free parameters; are the faint-end slope, normalization, 
and redshift evolution parameter.
Fig. 7 shows that our quasar space density at $z\sim5$ are higher than the result of Fontanot et al. (2007) and thus our results are not contradictory to their result, once the
redshift difference is corrected. Here it should be mentioned that Fontanot et al. (2007) adopted a different method in deriving the photometric completeness from
other surveys, that may introduce a possible systematic difference from other studies (see also Glikman et al. 2010). This effect is seen in Fig. 6, where the inferred bright-end quasar space density is different between the results by Fontanot et al. (2007) and by \cite{2006AJ....131.2766R} even although the same SDSS quasar sample is used in the two studies. This suggests that the completeness adopted in Fontanot et al. (2007) may be underestimated, and accordingly the
quasar space density is possibly overestimated by a factor of $\sim2-3$.
In the case that we derive the faint end of the QLF at $z\sim5$ by using the completeness which is 
calculated by Richards et al. (2006), the QLF of Fontanot et al. (2007) shifts toward lower space density in Fig. 7, i.e., well below our upper limits.
\begin{figure}
\begin{center}

\includegraphics[bb=0 15 600 540,clip,width=8.5cm]{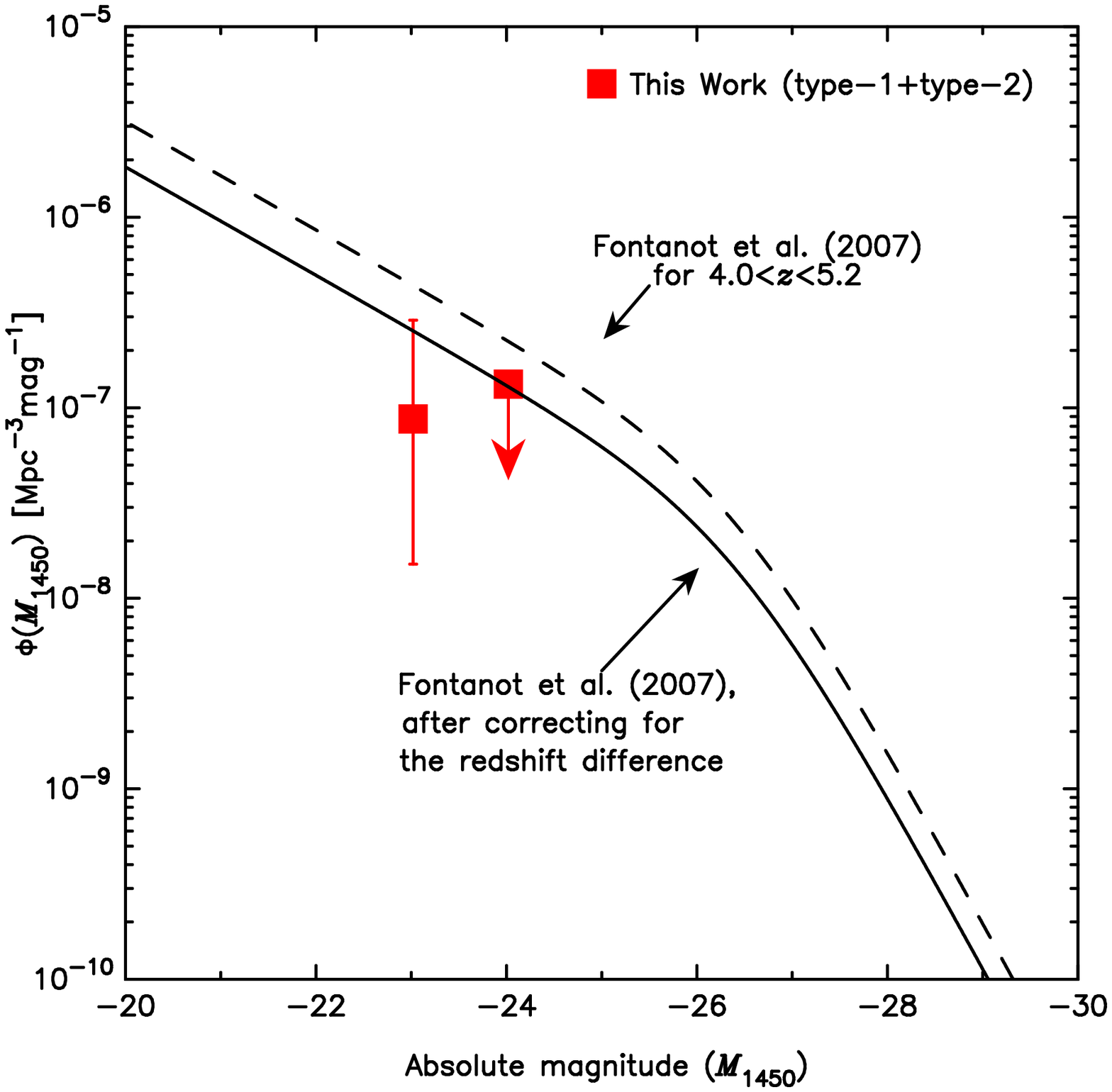}

\label{fig:quasar luminosity function}
\caption{Comparison of $z\sim5$ quasar luminosity functions when we include a type-2 quasar at $z\sim5.07$. The red squares show our results when we include a type-2 quasar at $z\sim5.07$. The black dashed and solid lines are the QLF  reported by Fontanot et al. (2007), before and after correcting for the redshift difference between their study and ours (assuming pure density evolution, see the main text for details).}

\end{center}

\end{figure}


In order to constrain the faint end of the QLF at $z\sim5$ quantitatively, we search for parameter values that satisfy our result. Here we adopt the following double power-law function:
\begin{eqnarray}
\Phi(M_{1450}, z) = \hspace{5.5cm} \nonumber \\
 \frac{\Phi(M^*_{1450})}{10^{0.4(\alpha+1)(M_{1450}-M^*_{1450})}+10^{0.4(\beta+1)(M_{1450}-M^*_{1450})}},
\end{eqnarray}
where $\alpha$, $\beta$, $\Phi(M^{*}_{1450})$, and $M^*_{1450}$ are the bright-end slope, the faint-end slope, the normalization of the luminosity function, and the characteristic absolute magnitude, respectively. Among the four parameters, the bright-end slope ($\alpha$) is fixed to be $\alpha = -3.31$ based on the SDSS results (see Fontanot et al. 2007). The parameter ranges which satisfy our results are shown in Fig. 8. Note that it is important to examine the redshift evolution of the faint-end slope and break magnitude, because such parameters give us useful constraints on the evolutionary model of SMBHs and quasars \citep[e.g.,][]{2006ApJ...639..700H,2007ApJ...654..731H}. By taking into account of the results obtained in COSMOS for $z\sim5$, the break magnitude in the QLF is brighter than $M^*_{\rm 1450} \sim -26$ at $z\sim5$. This is significantly brighter than the QLF break magnitude for $z\sim4$ reported by Ikeda et al. (2011) and Glikman et al. (2011), as shown in Fig. 8. A possible explanation for this evolution is that the mass accretion in most quasars at $z\sim5$ is higher than at $z\sim4$ (although the quasar number density is lower at $z\sim5$), which makes the characteristic magnitude brighter at $z\sim5$ than at $z\sim4$.
\begin{figure}
\begin{center}
\includegraphics[bb=10 25 400 750,clip,width=5cm,angle=270]{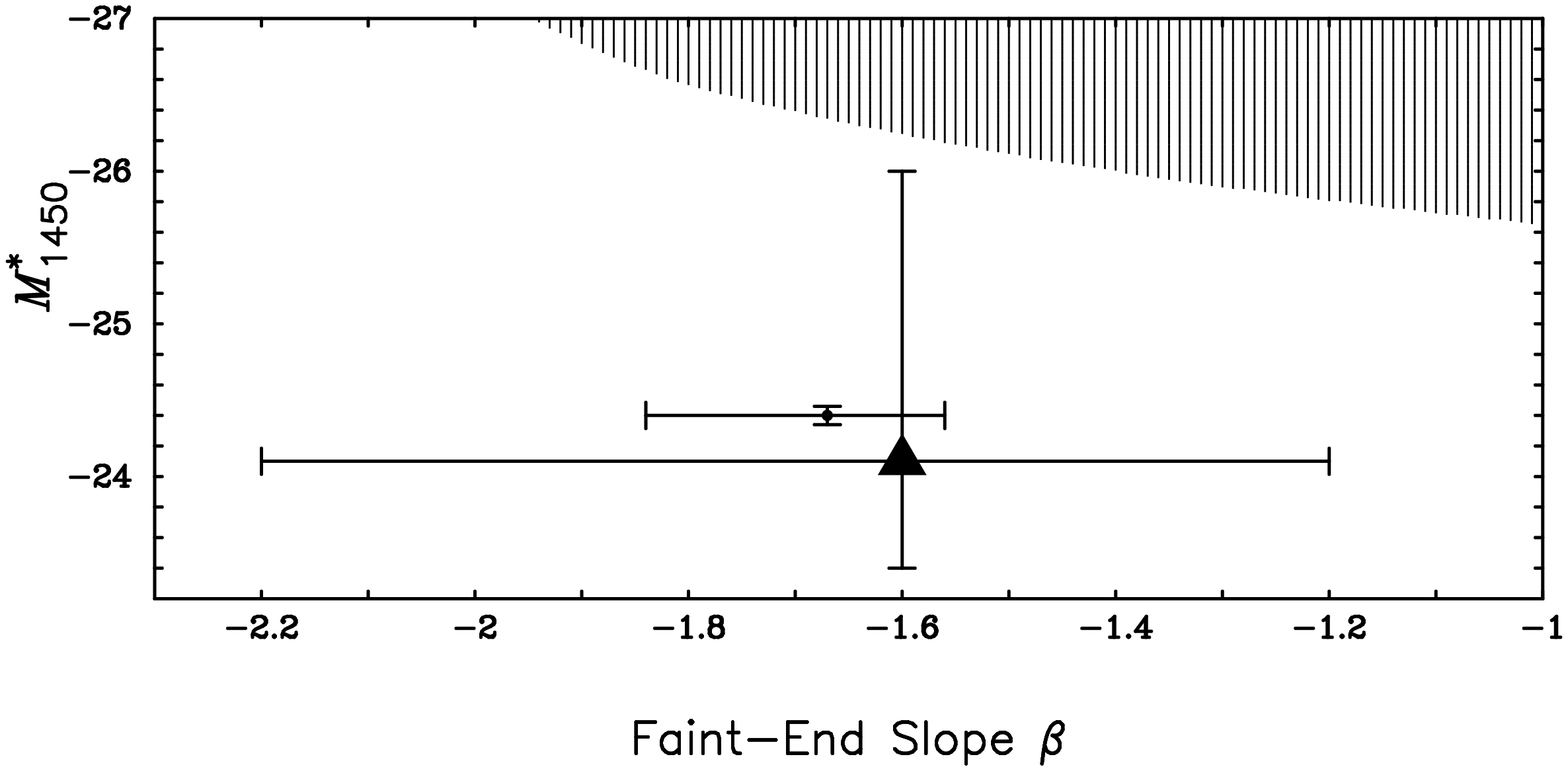}

\label{fig:quasar luminosity function}
\caption{Constraints on the faint-end slope $\beta$ and the break magnitude $M_{\rm 1450}^*$. The shaded region shows the parameter space that is consistent with the inferred
1$\sigma$ confidence upper limits on the quasar space density at $z\sim5$ derived in our study. A dot and triangle are the results inferred at
$z\sim4$ by Ikeda et al. (2011) and Glikman et al. (2011), respectively.
Note that we estimate the errors of $M^*_{\rm 1450}$ and the faint-end slope by applying a weighted least-squares method 
while Glikman et al. (2011) estimated them by applying the STY maximum-likelihood method \citep{1988MNRAS.232..431E}.
}

\end{center}
\end{figure}

Here we discuss the evolution of the quasar space density in the context of the AGN downsizing. 
The evolution of the quasar space density for different absolute magnitude bins provides important information 
to constrain the evolution of SMBHs. Therefore we plot the quasar space density for different absolute magnitude bins as a function of redshift in Fig. 9. 
Although there are a number of low-luminosity quasar surveys at  $z\sim3$ \citep{2003A&A...408..499W, 2004ApJ...605..625H, 2007A&A...461...39F, 2007A&A...472..443B}, we plot only the results of the 2dF-SDSS LRG and Quasar Survey (2SLAQ; Croom et al. 2009), the Spitzer Wide-area Infrared Extragalactic Legacy Survey \citep[SWIRE;][]{2008ApJ...675...49S}, and SDSS \citep{2006AJ....131.2766R}, in order to avoid data with large statistical errors. While most studies at $z < 3$ suggest consistent results (i.e., the AGN downsizing), the situation is rather controversial at $z > 3$. Recently, the new $z\sim4$ QLF results of the NOAO Deep Wide-Field Survey (NDWFS) and the Deep Lens Survey (DLS) are reported by Glikman et al. (2011). Interestingly, the results of Glikman et al. (2011) suggest constant or higher number densities of low-luminosity QSOs at $z>3$, 
while our COSMOS result suggests a continuous decrease of these objects from $z\sim2$ to $z\sim5$.

Our result is consistent with the downsizing AGN evolution suggested by
previous quasar surveys at lower-$z$ both in the optical and X-ray (e.g., Croom et al. 2009; Ueda et al. 2003; Hasinger et al. 2005).
\cite{2010AJ....139..906W} reported the faint end of the QLF at $z\sim6$ although there is a large error bar for the faintest magnitude bin because only one quasar was found. 
The $z\sim6$ space density at $M_{\rm 1450}\sim-22$ is
lower than the upper limit of our $z\sim5$ space density at the same magnitude and this result is also consistent with the AGN downsizing evolution.
However, the results of Glikman et al. (2011) require a different picture at $z>3$,
being inconsistent with the  downsizing scenario. It is not obvious what is causing this discrepancy. 
Masters et al. (submitted to the ApJ) stated that cosmic variance cannot be 
responsible for the observed discrepancy 
in space density of low-luminosity quasars between 
the COSMOS and the DLS NDWFS fields.
If this discrepancy is due to the difference 
in the quasar selection criteria, then this discrepancy 
could be caused by the presence or absence of the point 
source selection on the HST images.
However, both Ikeda et al. (2011) and Glikman et al. (2011) obtained the spectra of most quasar candidates to remove the contaminations.
Consequently, we cannot explain this discrepancy due to the difference
of the quasar selection criteria.
Therefore it remains important that we search for 
low-luminosity quasars at high redshift in other fields and 
derive the faint end of the QLF.
We also plot the quasar space density measured in the GOODS fields (Fontanot et al. 2007) in Fig. 9, 
which is consistent with both results from Glikman et al. (2011) and COSMOS due to its large uncertainty.
 Further observations of low-luminosity quasars in a wider survey area are crucial to derive firm constraints 
 on the scenarios of the quasar evolution, especially at $z>4$.
 \begin{figure}
\begin{center}
\includegraphics[bb=0 0 550 550,clip,width=8cm]{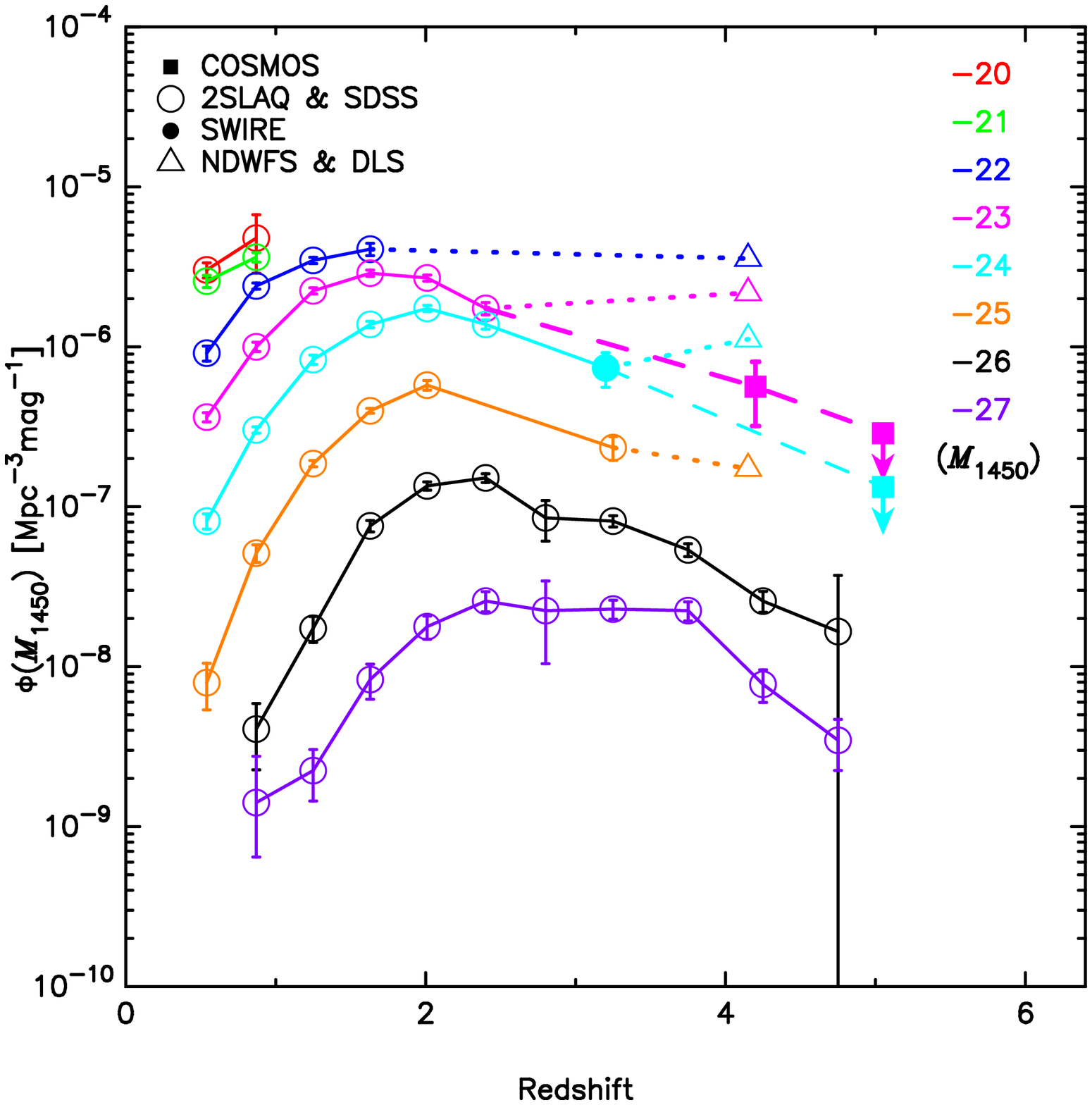}

\end{center}
\caption{The redshift evolution of the quasar space density. Red, green, blue, magenta, light blue, orange, black, purple lines are the space density of quasars with $M_{1450} = -20, -21, -22, -23, -24, -25, -26$, \rm and $-27$, respectively. Filled squares, open circles, filled circle, and open triangles denote the results of different quasar
surveys, as described in the upper-left side of the panel (see the main text for details).
For the comparison, the results of COSMOS and previous
surveys are connected by the dashed lines while the
results of NDWFS $\&$ DLS (Glikman et al. 2011) are
connected by the dotted lines.}

\label{fig:Redshift evolution of the quasar space density}

\end{figure}
\\
 
\section{Summary}
In order to examine the faint end of the QLF at $z\sim5$, we select photometric candidates of quasars at $z\sim5$ in the COSMOS field. The main results of this study are:
\begin{itemize}
\item Although we discover the type-2 quasar at $z\sim5.07$, no type-1 quasars at $z\sim5$ are identified through the follow-up spectroscopic observation.

\item The upper limits on the type-1 quasar space density are $\Phi<$ 1.33 $\times$ 10$^{-7}$ Mpc$^{-3}$ mag$^{-1}$ for 
$-24.52 < M_{1450} < -23.52$ and  $\Phi<$ 2.88 $\times$ 10$^{-7}$ Mpc$^{-3}$ mag$^{-1}$ for $-23.52 < M_{1450} < -22.52$.

\item The quasar space density and its error when we include a type-2 quasar at $z\sim5.07$ are  $\Phi=$ $0.87_{-0.72}^{+2.01}$ $\times$ 10$^{-7}$ Mpc$^{-3}$ mag$^{-1}$ for $-23.52 < M_{1450} < -22.52$.

\item The derived upper limits on the quasar space density are consistent with the QLF inferred by the previous works at $z\sim5$.

\item The characteristic absolute magnitude of the QLF shows a significant redshift evolution between $z\sim4$ ($M^{*}_{1450}>-26$) and $z\sim5$ ($M^{*}_{1450}<-26$).

\item A continuous decrease of the space density of low-luminosity ($-24\lesssim M_{1450}\lesssim -23$) quasars is inferred, that is roughly consistent with the picture of the AGN downsizing evolution.
\end{itemize}

\acknowledgments
We would like to thank the Subaru staffs for their invaluable help and all members of the COSMOS team. This work was financially supported in part by the Japan Society for the Promotion of Science (JSPS; Grant Nos. 23244031 and 23654068). This work was partly supported also by the FIRST program ``Subaru Measurements of Images and Redshifts (SuMIRe)'', that is initiated by the Council for Science and Technology Policy (CSTP). KM is financially supported by the JSPS through the JSPS Research Fellowship.

\bibliography{adssample}
\end{document}